\documentclass[twoside,twocolumn,10pt]{article}
\usepackage[pdftex]{color}
\usepackage{wscg}
\usepackage{url}
\usepackage[english]{babel}
\usepackage{graphicx}
\graphicspath{{./figures/}}
\usepackage{caption}
\DeclareCaptionType{copyrightbox}
\usepackage{tabu,booktabs}
\usepackage[sort&compress,numbers]{natbib}
\usepackage{microtype}
\ifdefined\anonymous
	\usepackage[switch]{lineno}
	\linenumbers
\else
	\usepackage{nopageno}
	\usepackage{flushend}
\fi
\usepackage[pdfusetitle]{hyperref}

\makeatletter
\def\affiliation#1{\def\@affiliation{#1}}
\def\@maketitle{%
  \begin{center}%
  \let \footnote \thanks
    \sffamily
    {\fontsize{16pt}{19.2pt} \bfseries \@title \par}%
    \vskip 1.0em%
    {%
      \lineskip .5em%
      \begin{tabular}[t]{c}%
        \@author
      \end{tabular}%
      \vskip 0.5em%
      \@affiliation%
      \par}%
  \end{center}%
  \par
  \vskip 0.5em}
\makeatother

\ifdefined\anonymous
	\author{Author 1}
	\affiliation{Faculty of ...\\University of ...\\Address ...\\Country ...\\author1@domain}
\else
	\author{David Barina}
	\affiliation{Centre of Excellence IT4Innovations\\Faculty of Information Technology\\Brno University of Technology\\Bozetechova 1/2, Brno\\Czech Republic\\ibarina@fit.vutbr.cz}
\fi
\title{Comparison of Lossless Image Formats}

\makeatletter
\def\Uslash{\mathbin{\mathchar`\/}\@ifnextchar{/}{\kern-.15em}{}}
\g@addto@macro\UrlSpecials{\do \/ {\Uslash}}
\def\Ucolon{\mathbin{\mathchar`:}\@ifnextchar{/}{\kern-.1em}{}}
\g@addto@macro\UrlSpecials{\do : {\Ucolon}}
\makeatother

\begin{document}

\twocolumn[{\csname @twocolumnfalse\endcsname

\maketitle

\begin{abstract}
In recent years, a bag with image and video compression formats has been torn.
However, most of them are focused on lossy compression and only marginally support the lossless mode.
In this paper, I will focus on lossless formats and the critical question: "Which one is the most efficient?"
It turned out that FLIF is currently the most efficient format for lossless image compression.
This finding is in contrast to that FLIF developers stopped its development in favor of JPEG XL.
\end{abstract}

\subsection*{Keywords}
Lossless Image Compression, PNG, JPEG-LS, JPEG 2000, JPEG XR, WebP, WebP 2, H.265, FLIF, AVIF, JPEG XL

\vspace*{1.0\baselineskip}
}]

\section{Introduction}
\label{sec:Introduction}

\copyrightspace

This paper compares the compression performance of state-of-the-art formats for lossless image compression.
It deals with the essential formats from old PNG (1992) to modern JPEG XL (2020).
The comparison is performed on three datasets.
The first one represents high-resolution photos.
The second one consists of artificial images (illustrations).
And the third one is made up of scanned pages of books.
The motivation for our comparison is to answer the question of which format is most efficient in terms of bits per pixel.

For comparison, we use tools that find the most suitable parameterization of the given compression formats.
As far as we know, this makes us different from most published comparisons.
Under these conditions, it turns out that the most efficient format for all three datasets is FLIF, closely followed by JPEG XL.
This situation is specifically quite interesting because the FLIF development has stopped as JPEG XL supersedes it.

This paper consists of four sections.
The \nameref{sec:Introduction} section opens this paper.
The \nameref{sec:Background} section discusses the formats and datasets used.
The \nameref{sec:Results} section presents our measurements.
Finally, the \nameref{sec:Conclusions} section closes the paper.

\section{Background}
\label{sec:Background}

This section describes the individual lossless image formats in our comparison.
We dealt with the following formats: PNG, JPEG-LS, JPEG 2000, JPEG XR, WebP, WebP 2, H.265, FLIF, AVIF, and JPEG XL.

Lossless algorithms \cite{Karam2009} usually work on the principle of pixel prediction, the error of which is subject to entropic coding, or a dictionary compression method followed by an entropic encoder.
Some formats use a transformation instead of an explicit predictor to decorrelate pixels.
Then we talk about residue coding instead of prediction error.
However, the principle is the same.
The entropic encoder can be a simple (Golomb-Rice encoder) as well as a very specialized arithmetic encoder coupled with context modeling.
The paragraphs below describe the individual lossless compression methods in detail.

PNG (1995) \cite{Roelofs1999,Salomon2004} is a purely lossless format initially intended to replace the GIF format.
It is based on a predictor followed by the dictionary compression method DEFLATE (a combination of LZ77 and Huffman coding), which is usually implemented by the zlib library.
Several tools can test different predictors and different parameterizations of the DEFLATE method.
This will achieve an even better compression ratio than the zlib at the cost of several orders of magnitude slower compression.
The zopflipng tool is used in our comparison with the parameters \texttt{--iterations=500 --filters=01234mepb}.

\begin{figure*}
	\includegraphics[width=\linewidth]{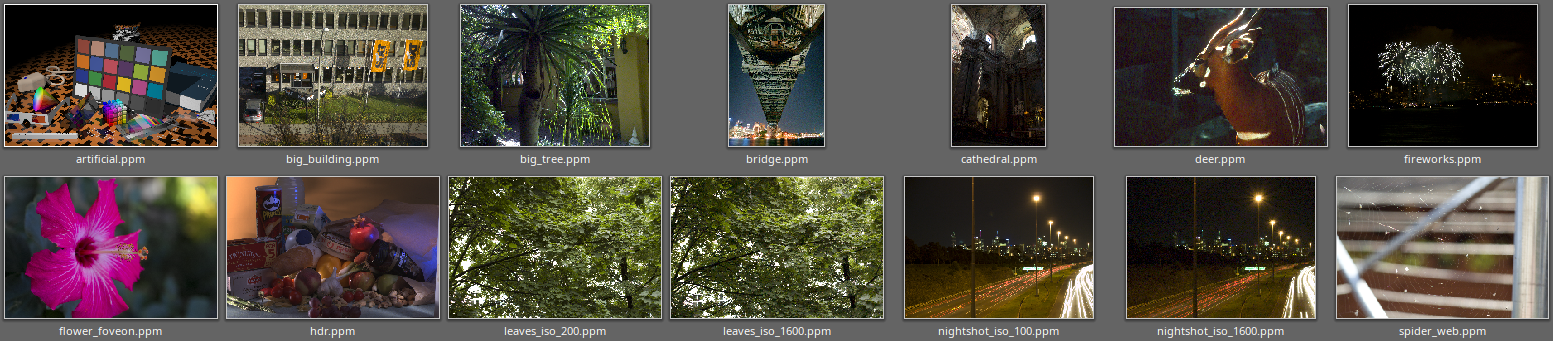}\medskip\\%
	\includegraphics[width=\linewidth]{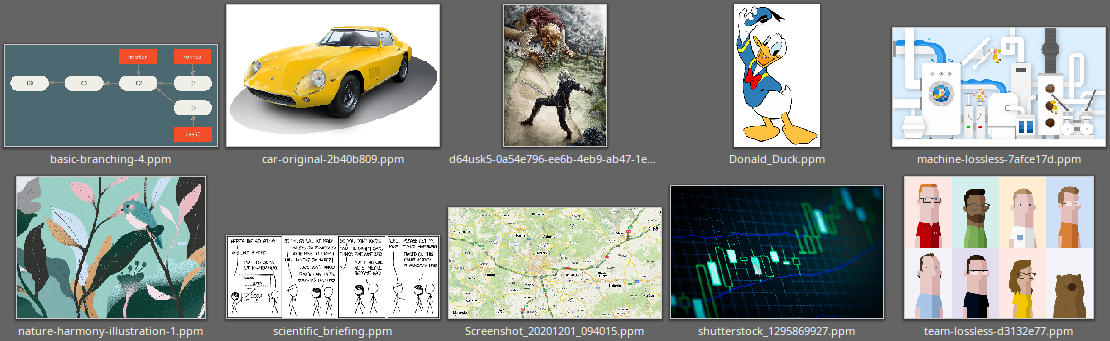}\medskip\\%
	\includegraphics[width=\linewidth]{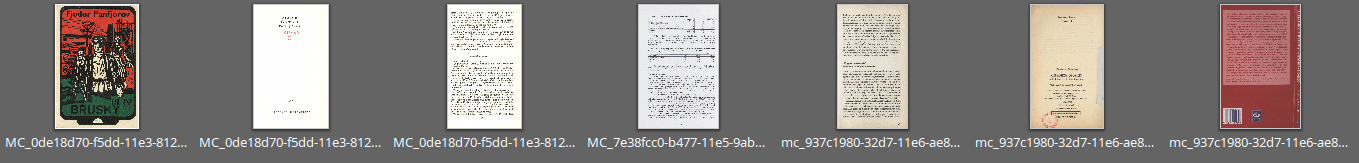}%
	\caption{Three datasets used in our comparison. From top: Photos (14 images), Illustrations (10 images), and Books (7 images).}
	\label{fig:dataset1}
\end{figure*}

JPEG-LS (1998) \cite{Weinberger2000} is a lossless image format created as a replacement for the lossless JPEG mode, which is inefficient.
This compression method consists of a predictor and a subsequent context entropic coder (Golomb-Rice coder).
The coder can also encode a sequence of the same pixels by a variant of the RLE method.
We use the libjpeg 1.58 library from Thomas Richter for compression with parameters \texttt{-ls 0 -c}.

JPEG 2000 \cite{Taubman2004} is a format based on a discrete wavelet transform \cite{Mallat2009} followed by a context arithmetic coder.
This makes it different from most other lossless compression methods, which are based on prediction error coding.
The format allows the choice of lossless or lossy compression paths.
Compression is incomparably faster than PNG.
There are several implementations of this format, of which we consider the Kakadu commercial library to be the best.
Kakadu version 8.0.5 is used for compression with parameters \texttt{Cblk=\{64,64\} Stiles=\{8192,8192\} Creversible=yes Clayers=1 Clevels=5}.

% \begin{figure*}
% 	\includegraphics[width=\linewidth]{dataset1}\medskip\\%
% 	\includegraphics[width=\linewidth]{dataset2}\medskip\\%
% 	\includegraphics[width=\linewidth]{dataset3}%
% 	\caption{Three datasets used in our comparison. From top: Photos (14 images), Illustrations (10 images), and Books (7 images).}
% 	\label{fig:dataset1}
% \end{figure*}

JPEG XR (HD Photo) \cite{Dufaux2009} is a 2009 format developed by Microsoft.
The format offers a lossless compression mode based on hierarchical discrete cosine transform (DCT) and Huffman coding.
As with JPEG 2000, the processing steps are the same for both lossless and lossy coding.
Reference software 1.41 with default parameters is used for compression.

WebP \cite{Google2020} is a 2010 compression format developed by Google.
The format allows the choice of lossy or lossless compression.
The lossless variant is based on a predictor followed by a combination of LZ77 and Huffman coding.
The cwebp reference tool with the \texttt{-lossless} parameter is used for compression.
WebP 2 is the successor of the WebP image format, currently in development.
The cwp2 reference tool with the \texttt{-effort 9 -q 100} parameters is used in our comparison.

H.265 (also HEVC) \cite{Sze2014} is a 2013 video compression format that supports lossless compression.
This format is again based on spatial prediction (without the use of DCT).
The format is a classic representative of hybrid video compression.
HEIC and BPG image formats are based on their Intra profile.
We use the x265 library version 3.4 for compression through the FFmpeg framework with parameters \texttt{-c:v libx265 -preset placebo -x265-params lossless=1}.

% https://ieeexplore.ieee.org/document/7532320
FLIF \cite{Sneyers2016,Soferman2015} is a lossless image format from 2015, strongly inspired by the FFV1 format.
The format is based on a predictor and a sophisticated arithmetic encoder MANIAC.
As with PNG, the compression ratio can be improved by searching for various parameters.
We use the flifcrush tool for this.
FLIF development has stopped since FLIF is superseded by JPEG XL.

% https://arxiv.org/pdf/2008.06091.pdf
AVIF is based on the AV1 \cite{Han2021} video format from 2018.
Also, this format is a classic representative of hybrid video compression.
The format offers both lossy and lossless compression.
The format uses a predictive-transform coding scheme for lossless mode, where the prediction comes from intra-frame reference pixels.
The residuals undergo a 2D transform, and then they are entropy coded using arithmetic coding.
In comparison, the libavif version 0.8.3 library with the \texttt{-l -c aom} parameters (libaom 2.0.0) was used.

JPEG XL (2020) \cite{Alakuijala2019,Alakuijala2020} is a new image format that supports both lossy and lossless compression.
ISO/IEC standard is still under development.
Its basic building blocks include transform, prediction, context modeling, and entropy coding (ANS \cite{Duda2015} or Huffman coding).
We use a reference software version 0.3.3 with \texttt{-q 100 -s 9} parameters.

% \begin{figure*}
% 	\includegraphics[width=\linewidth]{dataset1}\medskip\\%
% 	\includegraphics[width=\linewidth]{dataset2}\medskip\\%
% 	\includegraphics[width=\linewidth]{dataset3}%
% 	\caption{Three datasets used in our comparison.}
% 	\label{fig:dataset1}
% \end{figure*}

A dataset of 14 high-resolution photographic images \cite{Rawzor2015} in RGB24 format was used for comparison.
Thus, the uncompressed image has a bit rate of 24 bpp (bits per pixel).
The individual images are shown in Figure~\ref{fig:dataset1}.
The dataset occupies a total of 157 megapixels.
We will refer this dataset to it as Photos.
We also tested the compression performance on illustrations (artificial images, still 24 bpp).
This dataset occupies 10 megapixels.
This dataset was composed of images found on Google.
We will refer to it as Illustrations.
The last dataset used in our paper consists of seven scanned book's pages in 24 bpp and comprises 27 megapixels.
The scans were obtained from the collection of the Digital Library of the National Library of the CR.
This dataset will be called Books.

\section{Method and Results}
\label{sec:Results}

We use two quantities to evaluate the compression efficiency: a bitrate and compression ratio.
The bitrate is defined as the number of bits per single image pixel.
The compression ratio is defined as the ratio between the uncompressed and compressed file size.
The results are shown in Tables~\ref{tab:dataset1}, \ref{tab:dataset2}, and \ref{tab:dataset3}.

We evaluated the compression performance of all mentioned image formats on all three datasets.
It turned out that the clear winner is the FLIF format.
It is even more efficient than the newer JPEG XL.
Even though the development of FLIF was stopped in favor of JPEG XL.
It is worth noting that an exhaustive search found the FLIF format parameters.
JPEG XL has proven to be the second-best choice.
WebP 2 is also relatively efficient.

% The Impact of State-of-the-Art Techniques for Lossless Still Image Compression
The victory of the FLIF format is also in line with recent results of Rahman \textit{et al.}
In their paper \cite{Rahman2021}, the authors compared the compression performance of the lossless JPEG, JPEG 2000, PNG, JPEG-LS, JPEG XR, CALIC, HEIC, AVIF, WebP, and FLIF formats.
They concluded that the FLIF is optimal for all types of 24 bpp images, which they evaluated.

% Photos (dataset1, rgb8bit)
\begin{table}[t!]
	\begin{tabu} to \linewidth {X[l]|X[r]|X[r]}
		\toprule
			Format & Bitrate [bpp] & Compression Ratio \\
		\midrule
			PNG       & 11.461131 & 2.094034 : 1 \\
			JPEG-LS   & 10.588847 & 2.266535 : 1 \\
			JPEG 2000 & 10.620645 & 2.259749 : 1 \\
			JPEG XR   & 11.575239 & 2.073391 : 1 \\
			WebP      & 10.619910 & 2.259906 : 1 \\
			H.265     & 12.460111 & 1.926146 : 1 \\
			\textbf{FLIF}      &  \textbf{9.318886} & \textbf{2.575415} : 1 \\
			AVIF      & 12.039541 & 1.993431 : 1 \\
			WebP 2    & 10.007292 & 2.398251 : 1 \\
			JPEG XL   &  9.433458 & 2.544135 : 1 \\
		\bottomrule
	\end{tabu}
	\caption{Compression performance on the Photos dataset. The best result in bold.}
	\label{tab:dataset1}
\end{table}

% Illustrations (dataset2)
\begin{table}[t!]
	\begin{tabu} to \linewidth {X[l]|X[r]|X[r]}
		\toprule
			Format & Bitrate [bpp] & Compression Ratio \\
		\midrule
			PNG       & 4.628656 & 5.185090 : 1 \\
			JPEG-LS   & 6.994191 & 3.431419 : 1 \\
			JPEG 2000 & 6.094012 & 3.938292 : 1 \\
			JPEG XR   & 8.097638 & 2.963827 : 1 \\
			WebP      & 4.294325 & 5.588771 : 1 \\
			H.265     & 7.158238 & 3.352780 : 1 \\
			\textbf{FLIF}      & \textbf{3.394439} & \textbf{7.070387} : 1 \\
			AVIF      & 8.198868 & 2.927233 : 1 \\
			WebP 2    & 3.621495 & 6.627097 : 1 \\
			JPEG XL   & 3.473148 & 6.910157 : 1 \\
		\bottomrule
	\end{tabu}
	\caption{Compression performance on the Illustrations dataset. The best result in bold.}
	\label{tab:dataset2}
\end{table}

% Books (dataset3)
\begin{table}[t!]
	\begin{tabu} to \linewidth {X[l]|X[r]|X[r]}
		\toprule
			Format & Bitrate [bpp] & Compression Ratio \\
		\midrule
			PNG       &  8.694597 & 2.760334 : 1 \\
			JPEG-LS   &  8.394430 & 2.859038 : 1 \\
			JPEG 2000 &  7.303011 & 3.286315 : 1 \\
			JPEG XR   &  8.989639 & 2.669740 : 1 \\
			WebP      &  7.451733 & 3.220727 : 1 \\
			H.265     & 10.326125 & 2.324201 : 1 \\
			\textbf{FLIF}      &  \textbf{6.084763} & \textbf{3.944278} : 1 \\
			AVIF      & 10.398287 & 2.308072 : 1 \\
			WebP 2    &  6.890983 & 3.482812 : 1 \\
			JPEG XL   &  6.216347 & 3.860788 : 1 \\
		\bottomrule
	\end{tabu}
	\caption{Compression performance on the Books dataset. The best result in bold.}
	\label{tab:dataset3}
	\vskip-1pt
\end{table}

Interestingly, all algorithms achieve a compression ratio about three times higher on the Illustrations dataset than on Photos and about two times higher than on Books.
It also says that the Photos dataset is the hardest to compress for lossless algorithms, whereas the Illustrations are the easiest task.

We thought even compare the computing needs of the individual formats.
However, such a comparison would not be fair for two reasons.
Firstly, for some formats, we use an exhaustive search for suitable parameters; for others, we do not.
Secondly, compression and decompression ran in parallel on different machines with different hardware.

\section{Conclusions}
\label{sec:Conclusions}

We compared the compression efficiency of state-of-the-art formats for lossless image compression.
The comparison was performed on three different datasets.
It has been found that the most efficient lossless compression method is the FLIF format.
And this despite the fact that its development has stopped since JPEG XL supersedes FLIF.
Furthermore, it turned out that the JPEG XL has proven to be the second-best choice.
The third in line is the WebP 2 format.
It should be noted that some formats are still under development.

\paragraph{Acknowledgements}
This work has been supported by
the Technology Agency of the Czech Republic (TA CR) project AI for Traffic and Industry Vision (no. TH04010144) and Progressive Image Processing Algorithms (no. TH04010394).

% \clearpage

\bibliographystyle{myabbrvnat}
\bibliography{sources}

\end{document}